# Coexistience of phononic six-fold, four-fold and three-fold excitations in ternary antimonide Zr$_3$Ni$_3$Sb$_4$

Mingmin Zhong[a], Ying Liu[b]*, Feng Zhou[a], Minquan Kuang[a], Tie Yang[a], Xiaotian Wang[a]*, and Gang Zhang[c]*

[a]School of Physical Science and Technology, Southwest University, Chongqing 400715, China;
[b]School of Materials Science and Engineering, Hebei University of Technology, Tianjin 300130, China
[c]Institute of High Performance Computing, Agency for Science, Technology and Research (A*STAR), 138632, Singapore.
*Corresponding author.
email addresses: ying_liu@hebut.edu.cn; xiaotianwang@swu.edu.cn; zhangg@ihpc.a-star.edu.sg

Abstract

*Three-, four-, and six-fold excitations have significantly extended the subjects of condensed matter physics. There is an urgent need for a realistic material that can have coexisting 3-, 4-, and 6-fold excitations. However, these materials are uncommon because these excitations in electronic systems are usually broken by spin-orbit coupling (SOC) and normally far from the Fermi level. Unlike the case in electronic systems, the phonon systems with negligible SOC effect, not constrained by the Pauli exclusion principle, provide a feasible platform to realize these excitations in a wide frequency range. Hence, in this work, we demonstrate by first-principle calculations and symmetry analysis that perfect 3-, 4-, 6-fold excitations appear in the phonon dispersion rather than the band structures of Zr$_3$Ni$_3$Sb$_4$, which is a well-known indirect-gap semiconductor with an Y$_3$Au$_3$Sb$_4$-type structure. This material features 3-fold quadratic contact triple-point phonon, 4-fold Dirac point phonon, and 6-fold point phonon. Moreover, these nodal-point phonons are very robust to uniform strain. Two obvious phonon surface arcs of the [001] plane are extended in the whole Brillouin zone, which will facilitate their detection in future experimental studies. The current work provides an ideal model to investigate the rich excitations in a single material.*

Condensed matter systems [1] can host Dirac [2], Weyl [3], and Majorana fermions [4]. Dirac and Weyl fermions are well-studied in many topological semimetals and topological metals [5-10] and their nodal points are 4- and 2-fold degenerated band crossing points, respectively. Condensed matter systems can also contain novel fermions without counterparts in high energy physics, such as 3-, 6-, and 8-fold excitations [11-15] due to fewer constraints placed by the space group symmetries. However, topological materials with unconventional 6- and 8-fold fermionic excitations under crystalline symmetry are rarely studied. In 2019, Schröter et al. [16] proposed that AlPt is a 6-fold nodal-point topological semimetal and these 6-fold fermions can be regarded as a higher spin extension of Weyl fermions. Such an unconventional excitation was also predicted in $PdSb_2$ [15], $PtBi_2$ [17], and $Li_{12}Mg_3Si_4$ [18]. In 2020, Zhang predicted that 8-fold fermionic excitations appeared in high quality $TaTe_4$ single crystals [14]. Unfortunately, these candidates with 6- and 8-fold fermionic excitations may have their own disadvantages. The 6-fold fermionic excitations in $Li_{12}Mg_3Si_4$ [18] will be influenced by spin-orbit coupling (SOC), the unconventional quasiparticle excitations in $TaTe_4$ and AlPt [14,16] are far from the Fermi level, and the $PdSb_2$ [15] hosts many trivial Fermi surfaces.

Topological bosons [19-35], have been intensively studied, are also elementary excitations in condensed matter systems. Phonons are the basic emergent boson of crystalline lattice, and similar to fermionic electrons, topological phonons also exist in crystalline solids due to the crystal symmetry constraints. So far, 2-fold Weyl phonons [23], along with 3-fold nodal point phonons and 4-fold Dirac phonons [24,25], have been proposed in phonon systems. However, to the best of our knowledge, 6-fold degenerated phonons, which are maximal degenerate in phonon systems, have not been predicted. In this work, first-principle calculations and symmetry analysis were used to study the topological phonons in a Zintl phase $Zr_3Ni_3Sb_4$ compound that was prepared by Wang et al. [36] via arc-melting with an $Y_3Au_3Sb_4$-type structure. The structural models of a $Zr_3Ni_3Sb_4$ unit cell and a primitive cell are shown in Figure 1(a) and Figure 1(b), respectively. The Zr, Ni, and Sb are located at 12a (3/8, 0, 1/4), 12b (7/8, 0, 1/4), and 16c (0.08207, 0.08207, 0.08207), respectively. It is well-known that $Zr_3Ni_3Sb_4$ is an indirect semiconductor as shown by the electronic structures in Figure 1c. This compound has been widely used as a parent compound to fabricate a pair of n-type and p-type thermoelectric materials [37]. Here, we propose that $Zr_3Ni_3Sb_4$ hosts three kinds of phonon excitations, namely, 6-fold nodal point phonon at H point, 3-fold nodal point at Γ point, and 4-fold Dirac point at P point. A detailed $k \cdot p$ model has also been constructed for these three points. The phonon surface states of the [001] plane are presented to demonstrate to show the Fermi arc states in $Zr_3Ni_3Sb_4$ and we also examine the dynamic stability and topological signatures under uniform strains.

Density functional theory [38] was used to calculate the electronic structures and we used the GGA-PBE [39] formalism for the exchange-correlation functional. The projector augmented-wave [40] method is used for the interactions between ions and valence electrons and the energy cutoff was set to 500 eV. A Γ-centered $k$-mesh of $10 \times 10 \times 10$ was used for Brillouin zone (BZ) sampling. The lattice dynamic calculations are performed to obtain the phonon dispersion of $Zr_3Ni_3Sb_4$ at equilibrium and strained lattice constants in the PHONOPY package [41] using density functional perturbation theory [42]. The topological behaviors of the [001] phonon surface states are calculated by constructing a Wannier tight-binding Hamiltonian of phonons [43].

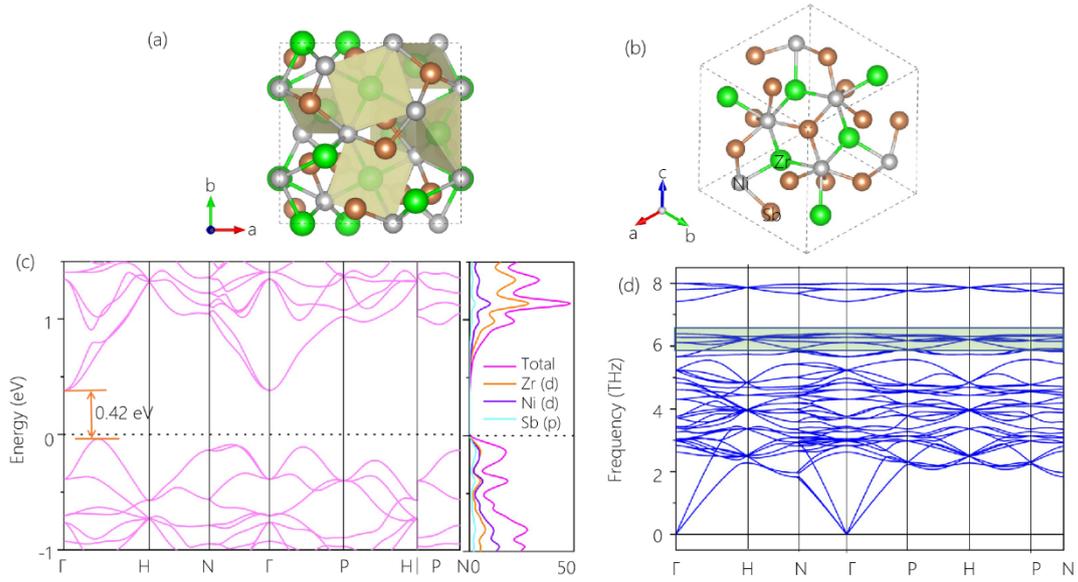

Figure 1. The unit cell (a) and primate cell (b) of $Zr_3Ni_3Sb_4$. The Zr, Ni, Sb atoms are shown by green, grey, and brown balls, respectively. The band structures and the PDOS of $Zr_3Ni_3Sb_4$ compound are shown in (c) and the calculated phonon dispersion of $Zr_3Ni_3Sb_4$ is shown in (d).

The structure of $Zr_3Ni_3Sb_4$ was totally relaxed and the calculated lattice constants were found to be a = b = c = 9.1346 Å, which agrees well with the experimental values of a = b = c = 9.066 Å [36]. The primate cell of $Zr_3Ni_3Sb_4$ shown in Figure 1(b) was used to calculate the band structure. Cubic type $Zr_3Ni_3Sb_4$ belongs to space group $I\bar{4}3d$ with a space group (SG) number of 220. The 3D BZ is shown in Figure 2(a) and the band structures and the density of states are shown in Figure 1(c). Obviously, $Zr_3Ni_3Sb_4$ is a semiconductor with an indirect band gap of 0.42 eV.

Figure 1(d) shows the calculated phonon spectra of $Zr_3Ni_3Sb_4$ along the high symmetry points $\Gamma - H - N - \Gamma - P - H - P - N$ as depicted in Figure 2(a). It can be seen that no acoustic-optical branch gap appears in the phonon spectra and $Zr_3Ni_3Sb_4$ is dynamically stable due to the absence of imaginary frequencies. We observe rich types of nodal point-phonons in the 6.10-6.25 THz region, as shown in Figure 2(b). A series of phonon band crossing points appears in this region as a 6-fold point (SFP) at H, a quadratic contact triple point (QCTP) at $\Gamma$, and a 4-fold Dirac point (DP) at P and because these are phonons we do not need to consider the Pauli exclusion principle as needed in electronic systems.

Therefore, these rich 3-, 4-, and 6-fold phonons can be realized in a wide range of phonon frequencies. To our best knowledge, the QCTP-phonons and the SFP-phonons have not been considered to date.

We have studied these excitations individually. As shown in Figure 2(c), the DP at the P point with a frequency of ~6.18 THz is formed by a doubly degenerated phonon band and two non-degenerated phonon bands. These DP in Figure 2 (c) with a graphene-like linear phonon band dispersion is belonging to type I nodal points as described in reference [44]. The topological phonons at the P point can be confirmed on the basis of an effective model around P where there exists four independent operations, namely, $\{C_{3,\bar{1}\bar{1}1}^+|0\frac{1}{2}\frac{1}{2}\}$, $\{C_{2y}|0\frac{1}{2}\frac{1}{2}\}$, $\{C_{2x}|\frac{3}{2}\frac{3}{2}0\}$ and $\{S_{4x}^+|\frac{1}{2}11\}$. The effective Hamiltonian can be written in a general form,

$$H_{DP}(\boldsymbol{k}) = \begin{bmatrix} h_{11}(\boldsymbol{k}) & h_{12}(\boldsymbol{k}) \\ h_{21}(\boldsymbol{k}) & h_{22}(\boldsymbol{k}) \end{bmatrix}, \quad (1)$$

where each $h_{ij}(\boldsymbol{k})$ is a $2 \times 2$ matrix as described in equations 2 and 3.

$$h_{11}(\boldsymbol{k}) = -h_{22}(\boldsymbol{k}) = \begin{bmatrix} 0 & vk_- \\ vk_+ & 0 \end{bmatrix}, \quad (2)$$

$$h_{12}(\boldsymbol{k}) = h_{21}^*(\boldsymbol{k}) = \begin{bmatrix} \beta k_z & \alpha k_+ \\ \alpha^* k_- & -\beta k_z \end{bmatrix}, \quad (3)$$

Here, $v$ is real while $\alpha$ and $\beta$ are imaginary parameters.

It can be seen that a degenerate point features a linear dispersion in any direction in *k*-space and since the P point is invariant under mirror $\widetilde{M}_{110}$ it carries a zero Chern number.

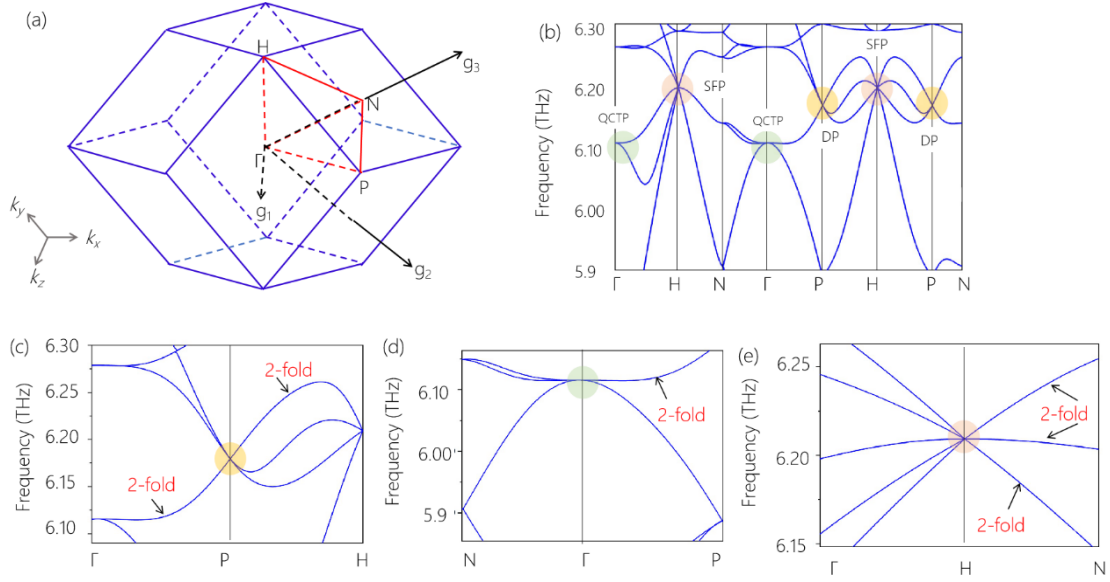

Figure 2. (a) Three-dimensional coordinate system of the BZ and some high symmetry points. (b) Enlarged frequencies of $Zr_3Ni_3Sb_4$ with rich nodal point-phonons. The SFP, QCTP, and DP, at H, Γ, and P, are highlighted by red, green, and yellow. The phonon band crossing points at the P, Γ, and H points, respectively (c)-(e).

Figure 2(d) shows the QCTP-phonon at the Γ point with a frequency of ~ 6.11 THz. The phonon bands at Γ exhibit a quadratic dispersion [45] along N-Γ-P paths as opposed to the linear dispersion at P. There is one non-degenerated phonon band and one 2-fold degenerated phonon band along the Γ-P direction. However, this band splits into two single ones along the N-Γ path. Although the 3-fold degenerate nodal point-fermions are widely proposed in electronic systems [46-50], such as, $APd_3$ (A = Pb, Sn) [46], nonsymmorphic $Ba_3Si_4$ [49], $Li_2NaN$ [50], and $TiB_2$ [48], the SOC will usually induce a gap to the band-crossing point and thus break the 3-fold nodal point-fermions. The QCTP in phononic system can be viewed as an ideal 3-fold nodal point because the phonon systems with negligible SOC effect. To capture the physics of the QCTP at the Γ point, an effective Hamiltonian may be written as

$$H_{QCTP}(k) = \begin{bmatrix} A_1 k_x^2 + A_2(k_y^2 + k_z^2) & Bk_xk_y & Bk_xk_z \\ Bk_xk_y & A_1 k_y^2 + A_2(k_x^2 + k_z^2) & Bk_yk_z \\ Bk_xk_z & Bk_yk_z & A_1 k_z^2 + A_2(k_y^2 + k_x^2) \end{bmatrix}. \quad (4)$$

Such a QCTP has zero topological charge due to the glide mirror and such a QCTP is a typical topological state that only occurs in spinless systems.

The 6-fold nodal point at the H point has a frequency of ~6.207 THz. As shown in Figure 2(e), three doubly degenerate phonon bands merge at the H point along the H-N direction to form a 6-fold phonon band crossing with linear dispersion. The 6-fold degenerate nodal point-phonons are the maximum degeneracy in phonon systems and these excitations in spin-free systems are not well studied. This work shows a realistic material to study the unconventional 6-fold nodal point-phonons for the first time. We construct a Hamiltonian of the phonon system around the H point to uncover the physical origin of the 6-fold excitations. The small group around the H point belongs to SG 220, which is generated by a reflection fourfold rotation ($S_{4x}$), a glide mirror ($\widetilde{M}_{110}$), and a threefold rotation. Generally, the effective model can be written as

$$H(\mathbf{k}) = \begin{bmatrix} h_{11}(\mathbf{k}) & h_{12}(\mathbf{k}) \\ h_{21}(\mathbf{k}) & h_{22}(\mathbf{k}) \end{bmatrix}. \tag{5}$$

Here, $\mathbf{k}$ is measured from the H point and each $h_{ij}(\mathbf{k})$ is a $3 \times 3$ matrix as described below. In detail,

$$h_{11}(\mathbf{k}) = -h_{22}(\mathbf{k}) = \gamma \mathbf{k} \cdot \mathbf{S}, \tag{6}$$

where $\mathbf{S}$ is the angular momentum for a spin-1 excitation that satisfies

$$[S_i, S_j] = i\epsilon_{ijk} S_k. \tag{7}$$

There are two spin-1 excitations of opposite helicity in the diagonal positions and

$$h_{12}(\mathbf{k}) = h_{21}^*(\mathbf{k}) = \begin{bmatrix} 0 & -\gamma_1 k_x & \gamma_2 k_y \\ -\gamma_2 k_x & 0 & \gamma_1 k_z \\ \gamma_1 k_y & \gamma_2 k_z & 0 \end{bmatrix}. \tag{8}$$

If we consider the limitations $|\gamma| \gg |\gamma_1|$ and $|\gamma| \gg |\gamma_2|$, this sixfold degenerate point is a composition of two decoupled spin-1 excitations that carries zero topological charge. The degenerate point is still gapless when taking the off-diagonal terms into consideration and no topological phase transition happens. Therefore, the 6-fold degenerate point at that H point has a Chern number of zero.

A phonon tight-binding Hamiltonian as implemented in the WANNIERTOOLS package [43] was used to calculate the phonon surface states of $Zr_3Ni_3Sb_4$ to further our understanding of the topological features at these multifold-degenerate points. The results of [001] phonon surface states and the corresponding isofrequency surfaces for $Zr_3Ni_3Sb_4$ at E1 = 6.21 THz and E2 = 6.10 THz are presented in Figure 3. The SFP, DP, and QCTP are shown by red, yellow, and green circles, respectively, in Figure 3(a). There are indeed two phonon surface arcs [51,52] emanating from these nodal points. These two phonon surface states around the SFP, DP, and QCTP are marked by red, black, and yellow arrows, respectively. It is well-known that clear surface states can facilitate their detection in future experiments. These phonon surface states are extended in the whole BZ on the [001] surface as shown in Figures 3(b) and Figure 3(c). The reason may be that the QTCP at $\Gamma$ is located at the center of the BZ while the SFP and DP at H and P are at the corner of the BZ and these phonon surface states will connect these projections and extend over the whole BZ of the [001] side surface (See the black arrows in Figures 3(b) and 3(c)).

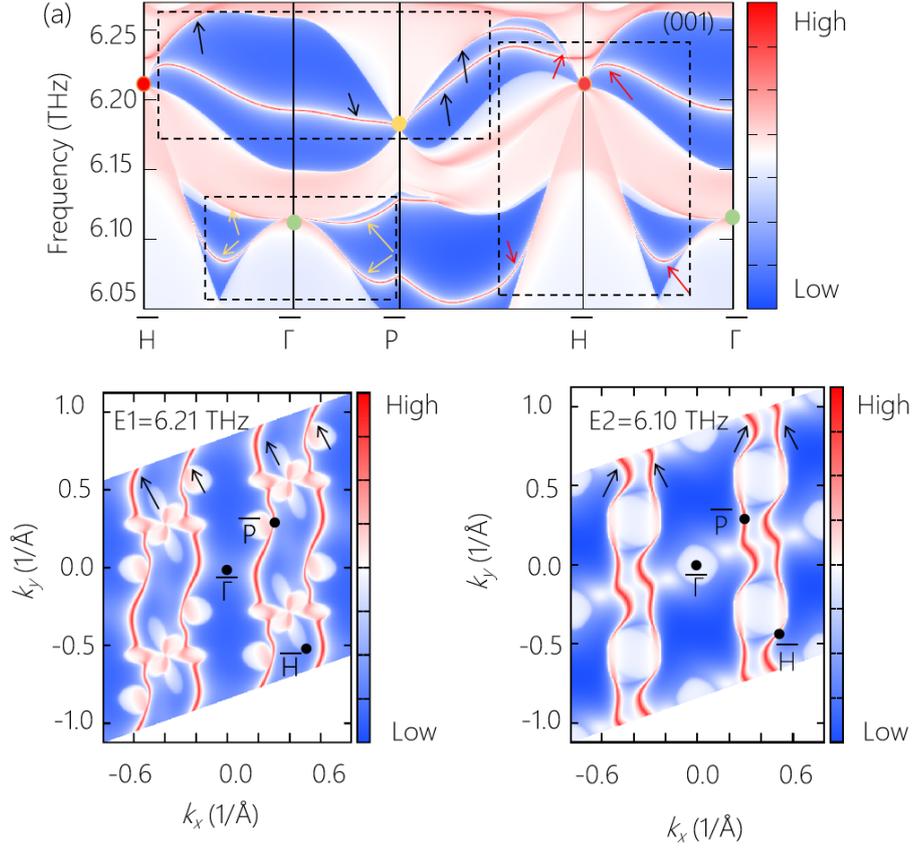

Figure 3. (a) The phonon surface states of the [001] plane. The SFP, DP, and QCTP are marked by red, yellow, and green circles. The phonon surface arc states around the SFP, DP, and QCTP are highlighted by red, black, and yellow arrows. The isofrequency surfaces at E1 = 6.21 THz and E2 = 6.10 THz are shown in (b) and (c), respectively.

These symmetry-protected excitations are very robust to uniform strains. Figures 4(a) and 4(b) show the calculated phonon dispersion of $Zr_3Ni_3Sb_4$ under ±5% uniform strains and it is dynamical stable in this range. The SFP, QCTP, and DP-phonons are still maintained in this region. The enlarged phonon dispersions of the highlighted parts of Figures 4(a) and 4(b) are shown in Figures 4(c) and 4(d), respectively. However, the regions of these multifold-degenerate points are reduced and increased by ~2 THz under the -5% and +5% uniform strains, respectively.

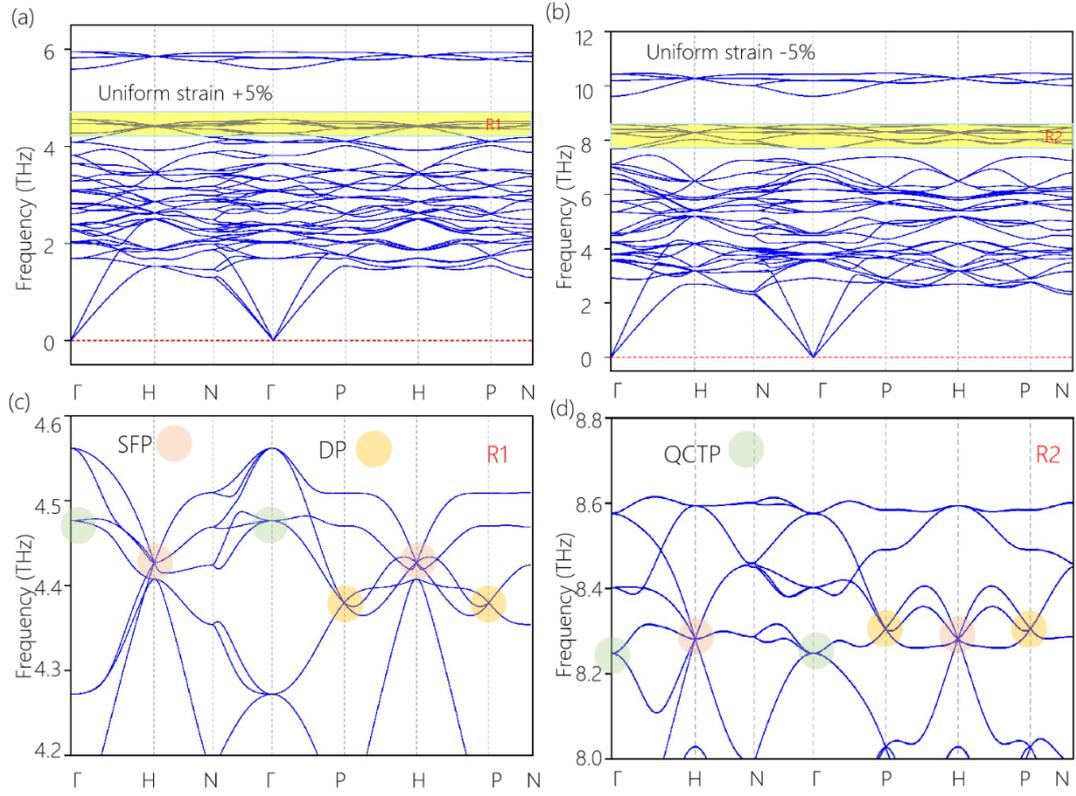

Figure 4. The phonon dispersion spectra of $Zr_3Ni_3Sb_4$ at (a) +5% uniform strain and (b) -5% uniform strains. Both (c) and (d) show the enlarged figures of (a) and (b). The SFP, QCTP, and DP are highlighted by different colored circles in (c) and (d).

In summary, we propose $Zr_3Ni_3Sb_4$ is an existing material with 3-, 4-, and 6-fold excitations. The SFP, DP, and QCTP-phonons at the H, P, and Γ points are in the frequency regions of 6.10-6.25 THz and $Zr_3Ni_3Sb_4$ can be viewed as an ideal candidate for investigating unconventional quasiparticles in bulk materials. Moreover, $Zr_3Ni_3Sb_4$ has phonon surface arcs that extend the entire BZ of the side surface. These multifold-degenerate points are maintained in phonon dispersion with ±5% uniform strains. Without SOC and the restriction of Pauli exclusion principle, $Zr_3Ni_3Sb_4$ is an ideal test case to study unconventional quasiparticles and we expect will be confirmed experimentally.